\documentclass[11pt]{article}
\pdfoutput=1
\usepackage{graphicx}
\usepackage{jheppub}

\usepackage{color}
\usepackage{amsmath}
\usepackage{pifont}
\usepackage{bbold}

\usepackage{bbm}
\usepackage{verbatim}   
\usepackage{subfigure}
\usepackage{acronym}

\usepackage{amsfonts}
\usepackage{amssymb}
\usepackage{mathrsfs}
\usepackage{multirow}
\usepackage{slashed}

\usepackage{cleveref}

\graphicspath{{./fig/}}

\def\bar#1{\overline{#1}}

\def\abs#1{\left| #1\right|}

\def\inv{^{\raise.15ex\hbox{${\scriptscriptstyle -}$}\kern-.05em 1}}
\def\lbar{{\lower.35ex\hbox{$\mathchar'26$}\mkern-10mu\lambda}} 

\let\<=\langle
\let\>=\rangle

\let\+=\uparrow

\def\T{\text{T}}
\renewcommand{\vec}[1]{\mathbf{#1}}

\begin{document}


%

\title{A boost for the EW SUSY hunt: monojet-like search for compressed sleptons at LHC14 with $100~\text{fb}^{-1}$}

\author[a]{Alan Barr,}
\emailAdd{A.Barr1@physics.ox.ac.uk}

\author[a,b]{James Scoville}
\emailAdd{james.scoville@physics.ox.ac.uk}

\affiliation[a]{Department of Physics, Denys Wilkinson Building, \\
Keble Road, Oxford OX1 3RH, UK}

\affiliation[b]{United States Air Force Institute of Technology,\\Wright-Patterson Air Force Base, OH 45433, USA}

\abstract{
Current Large Hadron Collider (LHC) analyses are blind to compressed supersymmetry (SUSY) models with sleptons near the lightest super partner (LSP) in mass: $m_{\tilde{l}} - m_{\tilde{\chi}_1^0} \equiv \Delta m \lesssim 60$ GeV.  We present a search sensitive to the very compressed range $3~\text{GeV} < \Delta m < 24~\text{GeV}$ using the channel $p p \rightarrow \tilde{l}^+ \tilde{l}^- +\rm{jet} \rightarrow$ $l^+ l^- \tilde{\chi}_1^0 \tilde{\chi}_1^0 +\rm{jet}$ with soft same-flavor leptons and one hard jet from initial state radiation ($p_{\T}^j >100$ GeV).  The sleptons recoil against the jet boosting them and their decay products, making the leptons detectable and providing substantial missing transverse momentum.  We use the kinematic variable $m_{\T 2}$ along with a different-flavor control region to reduce the large standard model backgrounds and control systematic uncertainty.  We find the analysis should allow LHC14 with $100~\text{fb}^{-1}$ to search for degenerate left-handed selectrons and smuons in the compressed region up to $m_{\tilde{l}_L} \lesssim 150$ GeV.  In addition, it should be sensitive to  $m_{\tilde{l}_L} \lesssim 110$ GeV for the very challenging case of auto-concealed SUSY, in which left-handed sleptons decay to the Kaluza-Klein tower of a modulino LSP which lives in $d=6$ extra dimensions.  In both the compressed spectrum and auto-concealed SUSY scenarios this analysis will need more data to improve on LEP2 limits for right-handed sleptons due to their smaller cross sections.
}

\maketitle


\section{Introduction}
\label{sec:intro}

Discovering supersymmetry (SUSY) at the Large Hadron Collider (LHC) requires distinguishing the signal decay products of SUSY particles (sparticles) from large standard model (SM) backgrounds.  In $R$-parity conserving SUSY in which the lightest super partner (LSP) is stable and invisible to the detector, methods commonly used to separate signal from backgrounds take advantage of the signal's large missing transverse momentum ($E_\T^{\rm{miss}}$), numerous jets from visible final state radiation, or highly energetic leptons \cite{deJong:2012zt}.  However, this becomes more challenging in the case of electroweak (EW) SUSY production where cross sections are modest, and is especially difficult for compressed EW SUSY scenarios in which the mass difference between the produced sparticle and the LSP is small.  Compressed SUSY scenarios result in less energetic final state radiation and less $E_\T^{\rm{miss}}$.  Not only are such processes harder to pick out of backgrounds during an analysis, but they may not even pass trigger requirements for the LHC detectors.  As a result, both CMS and ATLAS remain blind to compressed EW SUSY scenarios involving sleptons with $m_{\tilde{l}} - m_{\tilde{\chi}_1^0} \equiv \Delta m \lesssim 60$ GeV \cite{Aad:2014vma,Khachatryan:2014qwa}.

One way to circumvent the challenges of compressed SUSY is to search for events with an energetic jet from initial state radiation (ISR) \cite{Gunion:1999jr, Dreiner:2012gx, Mukhopadhyay:2014dsa, Han:2013usa}.  In such events the sparticles will recoil against the ISR, increasing both the energy of visible decay products and $E^{\rm{miss}}_\T$.  Such a strategy was used, for instance, by both ATLAS and CMS to search for top squarks with masses close to the LSP \cite{Aad:2014nra, CMS:2014yma}.  However monojet searches are typically not designed to find EW sparticles and therefore those analyses veto on leptons to suppress unwanted backgrounds.  Recently, several groups investigated the possibility of using a monojet, large $E_\T^{\rm{miss}}$, and soft leptons to pick out Higgsinos in a compressed spectrum \cite{Baer:2014kya, Han:2014kaa, Schwaller:2013baa}.  Motivated by their strategy and the current absence of slepton limits from the LHC for compressed models, we investigate the possibility of performing a similar search to discover sleptons.   

We rely on a hight-$p_\T$  ISR jet to boost pair produced same flavor opposite sign (SFOS) sleptons, which promptly decay to two neutralino LSPs and a lepton pair ($p p \rightarrow \tilde{l}^+ \tilde{l}^-j \rightarrow l^+ l^- \tilde{\chi}_1^0 \tilde{\chi}_1^0j$).  Like refs.~\cite{Baer:2014kya, Han:2014kaa} we search for a hard central jet, large $E_\T^{\rm{miss}}$, and a soft SFOS lepton pair.  However, unlike these studies we find it difficult to pick the signal out of the large backgrounds from leptonically decaying $t \bar{t}$ and $W^+W^-j$ using a veto on large $m_{ll}$ alone; since signal leptons come from opposite legs of the decay chains their angular separation is not always small.  Therefore we use the leptons' `stransverse' mass $m_{\T 2}$ \cite{Lester:1999tx, Barr:2003rg, Cheng:2008hk} to compress signal events into a narrow window and make the signal competitive with backgrounds.   

However, even this is insufficient to pick out the signal if there is $\mathcal{O}(20\%)$ systematic uncertainty in the remaining backgrounds.  Thus we use a different flavor lepton pair control region (i.e. $e \mu$) to subtract away these backgrounds, similar to the CMS study in ref.~\cite{CMS:2014jfa}. With this we are able to investigate the potential exclusion reach of the LHC at $\sqrt{s}=14$ TeV (LHC14) with $100~\text{fb}^{-1}$ if no excess in signal events are seen.  

In addition to the normal compressed spectrum scenario we also consider an auto-concealed (AC) SUSY scenario \cite{Dimopoulos:2014psa}.  In this extra-dimensional SUSY model the slepton can decay to a nearly continuous Kaluza-Klein (KK) tower of neutral states, with a denser number of states closer to the parent mass $m_{\tilde{l}}$.  The effect of the multiple states is phenomenological similar at colliders to a compressed spectrum of near-degenerate particles. The result is that current LHC searches are not sensitive to direct slepton production in auto-concealed scenarios \cite{Dimopoulos:2014psa}. \footnote{During the preparation of this paper, two searches were published which also attacked the compressed spectrum slepton problem.  The first to do so \cite{Dutta:2014jda} used vector boson fusion and missing transverse momentum to reduce standard model backgrounds, but required around $3000~\text{fb}^{-1}$ of data.  The second \cite{Han:2014aea} proposed a solution remarkably similar to ours, also using a high-$p_{\T}$ monojet to help generate $E_\T^{\rm{miss}}$ and employing a moving $m_{\T 2}$ window to distinguish signal from background.  However some of the details of signal selection differ and in particular they do not address background systematics as we do in this paper using the different flavor control region.}

\section{Simulation}
\label{sec:sim}

\subsection{Tools}
\label{sec:tools}

We investigated the prospect of discovering slepton production in near-degenerate scenarios at the LHC using simulations of $pp$ at $\sqrt{s}=14$ TeV, assuming an integrated luminosity of $100~\rm{fb}^{-1}$.
To generate signal events and SM backgrounds we used MadGraph5\_aMC@NLO at tree level \cite{Alwall:2014hca} paired with Pythia 6 \cite{Sjostrand:2006za} for showering and hadronization.  With the exception of taus, decays into leptons were done at the matrix element level (within Madgraph) thus retaining spin correlations and increasing generator efficiency.  Taus were decayed using Tauola \cite{Shekhovtsova:2012yq} within Pythia.  

For both the SUSY signal ($pp\rightarrow \tilde{l}^+\tilde{l}^-j$) and the dominant background ($pp \rightarrow t \bar{t}$) we used MLM matching with up to one additional jet (e.g. $pp\rightarrow \tilde{l}^+\tilde{l}^-jj$).  For all backgrounds we used a generator level cut $ R_{ll} \equiv \sqrt{\Delta \eta^2 + \Delta \phi ^ 2} >1.0$ between leptons and for all non-MLM matched processes we used a generator level cut on the leading jet $p_\T >80$ GeV to increase efficiency (both of which were looser than the final analysis level cuts and were checked to not impact the results).

Events were then fed into an analysis built by the authors within the program CheckMATE \cite{Drees:2013wra,Cacciari:2005hq,Read:2002hq}, which includes an improved ATLAS detector simulation in Delphes \cite{deFavereau:2013fsa}.  In our analysis, jets and isolated leptons were assigned as follows:

\begin{itemize}
\item Jets were defined using the anti-$k_\T$ algorithm \cite{Cacciari:2011ma,Cacciari:2008gp} with a distance parameter $0.4$, $\abs{\eta} < 4.5$, and $p_{\T} > 20$ GeV.  The assumed $b$-jet tagging efficiency is $80\%$.  Further details on the $b$-tagger can be found in ref.~\cite{Drees:2013wra}.
\item Isolated leptons were defined to mimic ref.~\cite{TheATLAScollaboration:2013hha}.  Electrons (muons) were required to have $\abs{\eta} < 2.47$, $p_{\T} > 7$ GeV and to not be within $R < 0.4$ of a reconstructed jet or $R < 0.1$ of another isolated lepton.  The $p_\T$ sum of tracks above 0.4(1) GeV within $R < 0.3$ was required to be less than $16(12)\%$ of the lepton $p_\T$.  Electrons had the additional requirement that the sum of energies within $R < 0.3$ was required to be less than $18\%$ of the electron energy.
\end{itemize}

\subsection{Slepton signal}
\label{sec:signal}

We consider two cases: a classic compressed SUSY scenario and an auto-concealed SUSY scenario, described in further detail below.  For the classic scenario we use a simplified model with degenerate selectrons and smuons, both left or right-handed, with all other sparticles decoupled.  The sleptons decay with a $100\%$ branching ratio to a lepton plus a bino-like LSP.

\begin{figure}[t]
  \centering
  \includegraphics[width=.7\columnwidth]{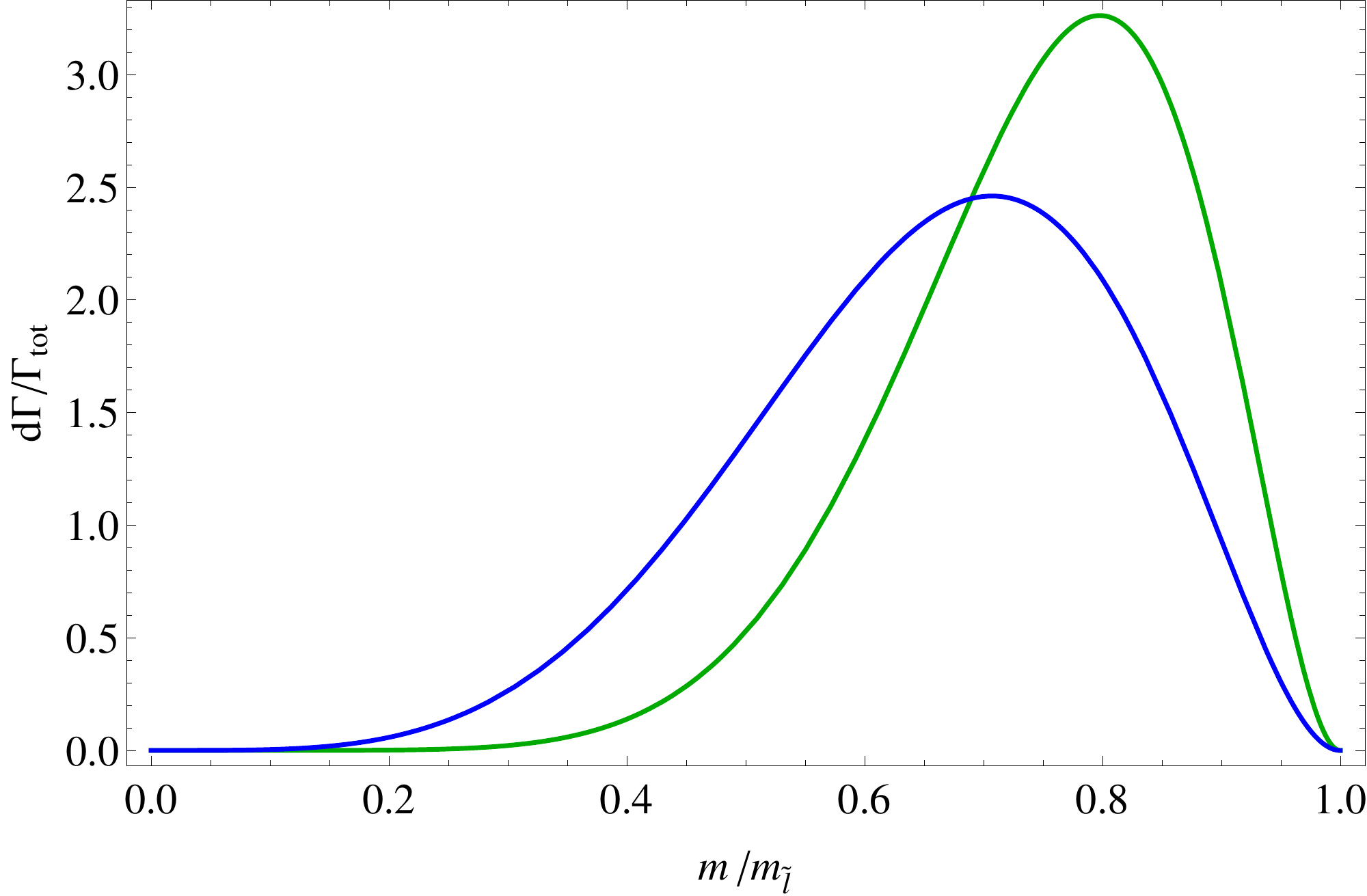}
  \caption{Differential decay rate for sleptons in an auto-concealed SUSY scenario.  Here sleptons decay to the KK-tower of a bulk modulino LSP living in $d=6$ (green) or $d=3$ (blue) extra dimensions.  The modulino mass is given as a ratio to the slepton mass $m_{\tilde{l}}$ on the horizontal axis.  Since the number of KK-modes grows like $\sim m^{d-1}$, the $d=6$ scenario results in a more compressed spectrum.}
  \label{fig:auto-concealed BR}
\end{figure}

In the auto-concealed SUSY scenario, sleptons are constrained to live on a brane in a $4+d$ dimensional bulk and can decay promptly to a lepton and a nearly continuous tower of bulk LSP KK-modes.  The branching ratio to KK-modes with mass $m$ is given by \cite{Dimopoulos:2014psa} \footnote{For decays to a modulino KK-tower.  For other LSP cases see \cite{Dimopoulos:2014psa}.}

\begin{equation}
\frac{d\Gamma}{\Gamma_\text{tot}}= \frac{1}{\Gamma_\text{tot}}\frac{\Omega_d}{(2 \pi)^d}\frac{ m_{\tilde{l}}^{d+3}}{8\pi M_*^{d+2}} \left(\frac{m}{m_{\tilde{l}}}\right)^{d+1}\left(1-\frac{m^2}{m_{\tilde{l}}^2}\right)^2 \frac{dm}{m_{\tilde{l}}}
\end{equation}
where $m_{\tilde{l}}$ is the parent slepton mass, $V$ is the bulk volume, $\Omega_d$ is the surface area of a $(d-1)$-sphere, $M_*$ is the fundamental gravitational scale, and $\Gamma_{\text{tot}}$ is the slepton total decay width.  Assuming the relevant scales are such that the slepton decay is prompt, this leads to a quasi-compressed spectrum as shown in Fig.~\ref{fig:auto-concealed BR}.

We approximate the continuous spectrum as in \cite{Dimopoulos:2014psa} by introducing $N=20$ new gauge neutral spin $1/2$ states in Pythia. The masses of these states $m_j$ fell into $N$ evenly spaced bins from 0 to the slepton mass $m_{\tilde{l}}$. The mass $m_j$ of the $j^{\rm{th}}$ state was given by the branching ratio-weighted average of masses in the $j^{\rm{th}}$ bin, and the branching fraction to this state was determined by the integrated width over the bin.

\subsection{Backgrounds}
\label{sec:bg}

We estimate SM backgrounds in the 2 leptons + 1 jet + $E_\T^{\rm{miss}}$ channel by modelling the processes below.  Tops and $W$'s are decayed leptonically within Madgraph. In all processes except $(Z/\gamma^*)j$, leptonic decays include taus (which are then decayed in Tauola); $(Z/\gamma^*)j$ decaying to taus is included in the $\tau \tau j$ background.  The backgrounds were:
\begin{enumerate}
\item $\tau^+ \tau^- j$.
\item $t \bar{t}$.
\item $W^+ W^- j$.
\item $l^+l^-(Z\rightarrow \nu \bar\nu)j$ where the leptons are primarily from $Z/\gamma^*$.  We will simply refer to this process as $ZZj$.
\item $l^+l^-W^\pm j$. Despite the fact that this process produces three visible leptons, its relatively large cross section combined with the possibility of losing a lepton makes this background more significant than $ZZj$.  We will refer to this process as $WZj$.
\item $tW$ with jets defined to come from light-flavor or $b$-quarks.  To increase generation efficiency it is required that the jet $p_{\T}$ be greater than 80 GeV.
\item $tq$ with jets defined to come from light-flavor or $b$-quarks and the same jet cut is imposed as in $tW$.
\item $l^+l^-j$ where the required missing transverse momentum comes primarily from mismeasured jets.  We will refer to this process as $Zj$.
\end{enumerate}
We also investigated the rare processes $t\bar{t}W$ and $t\bar{t}Z$ but found contributions from these processes was negligible.  

\section{Beating the backgrounds}
\label{sec: analysis}

\subsection{Key variables: $m_{\tau \tau}^2$ and $m_{\T 2}$}
\label{sec: variables}

The SM backgrounds discussed in Sec.~\ref{sec:bg} are much greater than the SUSY signal we are seeking.  To reduce the BG we use a series of kinematic cuts, including cuts on the reconstructed $m_{\tau \tau}^2$, which separates out the $\tau \tau j$ BG, and the `stransverse mass' $m_{\T 2}$, with which we define our final signal windows.  

To reconstruct the di-tau invariant mass squared $m_{\tau \tau}^2$ we follow ref.~\cite{Baer:2014kya} and use the fact that taus recoiling against a 100 GeV jet are highly relativistic and their decay products will be nearly parallel.  Thus in a fully leptonic di-tau decay we write the sum of the neutrinos' transverse momentums as 
\begin{equation}
\label{eq:pneutrinos}
\vec p_\T^{\rm{miss}} = \xi_1 \vec p_{\T}^{l_1} + \xi_2 \vec p_{\T}^{l_2}
\end{equation}
where $\vec p_\T^{\rm{miss}}$ is the missing transverse momentum vector and $\xi_n$ is a scale factor relating the transverse momentums of the $n^{\rm{th}}$ tau daughter neutrinos to the transverse momentum of the daughter electron or muon $\vec p_{\T}^{l_n}$.  Using this set of two equations we solve for the unknowns $\xi_1$ and $\xi_2$, with which we find the four-momenta of the taus: $p_{\tau_n} = (1+\xi_n) p_{l_n}$.  Then the di-tau invariant mass squared is
\begin{equation}
\label{eq:mtautau}
m_{\tau \tau}^2 = 2(1+\xi_1)(1+\xi_2)p_{l_1} \cdot p_{l_2}.
\end{equation}
This definition differs from ref.~\cite{Han:2014kaa} in that it allows for a negative invariant mass squared if $\xi < -1$ for a single $\xi$.\footnote{From eq.~(\ref{eq:pneutrinos}) we see this can occur for a missing transverse momentum vector nearly opposite to a lepton's $\vec p_{\T}$ and $p_\T^{miss}>p_\T^l$.  This can happen in $WWj$, for instance, when a neutrino and a lepton (possibly coming from different decay legs) are nearly back-to-back.}  In practice we find this definition slightly more effective at separating signal from background.

To determine the final signal regions we use the kinematic variable $m_{\T2}$, defined as 
\begin{equation}
m_{\T 2}(p_\T^{l_1},p_\T^{l_2};m_{\tilde{\chi}_1^0})=\underset{q_\T}{\min}\left[\max\left( m_\T(p_\T^{l_1},q_\T ; m_{\tilde{\chi}_1^0}), m_\T(p_\T^{l_2},p_\T^{miss} - q_\T ; m_{\tilde{\chi}_1^0}) \right)\right]
\end{equation}
where $m_{\tilde{\chi}_1^0}$ is the mass of the neutral particle which produces $E^{\rm{miss}}_\T$ and
\begin{equation}
m_\T(p_\T^l,q_\T;m_{\tilde{\chi}_1^0})= \sqrt{ m_l^2 + m_{\tilde{\chi}_1^0}^2 + 2 \left( E_\T^l E_\T^q-\vec{p}_\T \cdot  \vec{q}_\T \right) }.
\end{equation}

In a two-body semi-invisible decay, the stransverse mass variable $m_{\T2}$ provides an event-by-event bound in the space of masses of the parent and the invisible daughter particle. In practice, it is usually employed as a function which takes as an input the invisible particle's proposed mass, and returns the maximal lower bound on the mass of the parent particle. For example, in a pair-decay of sleptons of mass $m_{\tilde{l}}$ to leptons and LSPs of mass $m_{\tilde{\chi}_1^0}$ the function $m_{\T2}(m_{\tilde{\chi}_1^0})$ is always smaller than $m_{\tilde{l}}$ when the correct LSP mass is input. Since the same is not necessarily true for background processes, we can preferentially select the signal by imposing a requirement that $m_{\T2}(m_{\tilde{\chi}_1^0}) < m_{\tilde{l}}$ for our trial LSP and slepton masses. In this way, we can scan the $m_{\tilde l}$ -- $m_{\tilde{\chi}_1^0}$ plane. 

\subsection{Analysis cuts}
\label{sec:cuts}

Using the variables $m_{\tau \tau}^2$ and $m_{\T2}$ as defined above we use the following analysis cuts to pull the slepton signal out of the much larger SM backgrounds which also produce 2 leptons + 1 jet + $E_\T^{\rm{miss}}$: 

\begin{enumerate}
\item \label{step:no b} Veto events with a tagged $b$-jet to reduce backgrounds involving tops.

\item Require one hard central jet with $\abs{\eta} < 2.5$ and $p_{\T} > 100$ GeV.  Veto events with a second jet with $\abs{\eta} < 4.5$ and $p_\T>40$ GeV.

\item Require $E_\T^{\rm{miss}}$ of at least 100 GeV.  Also require $\abs{\Delta\phi}>1.5$ between the hard jet and $\vec{p}_\T^{miss}$.  This helps to reduce backgrounds from $Zj$ with $E_\T^{\rm{miss}}$ coming from mismeasured jet energy.

\item \label{step:2 lep} Require two SFOS isolated leptons with $R>1.3$ between them.  The large $R$ requirement significantly reduces backgrounds involving $Z/\gamma^*$ (in particular the otherwise large $WZj$ background in which one lepton escapes detection) and is more effective than a soft invariant mass veto since this would also veto signal leptons which tend to be soft.

\begin{figure}[t]
  \centering
  \includegraphics[width=.7\columnwidth]{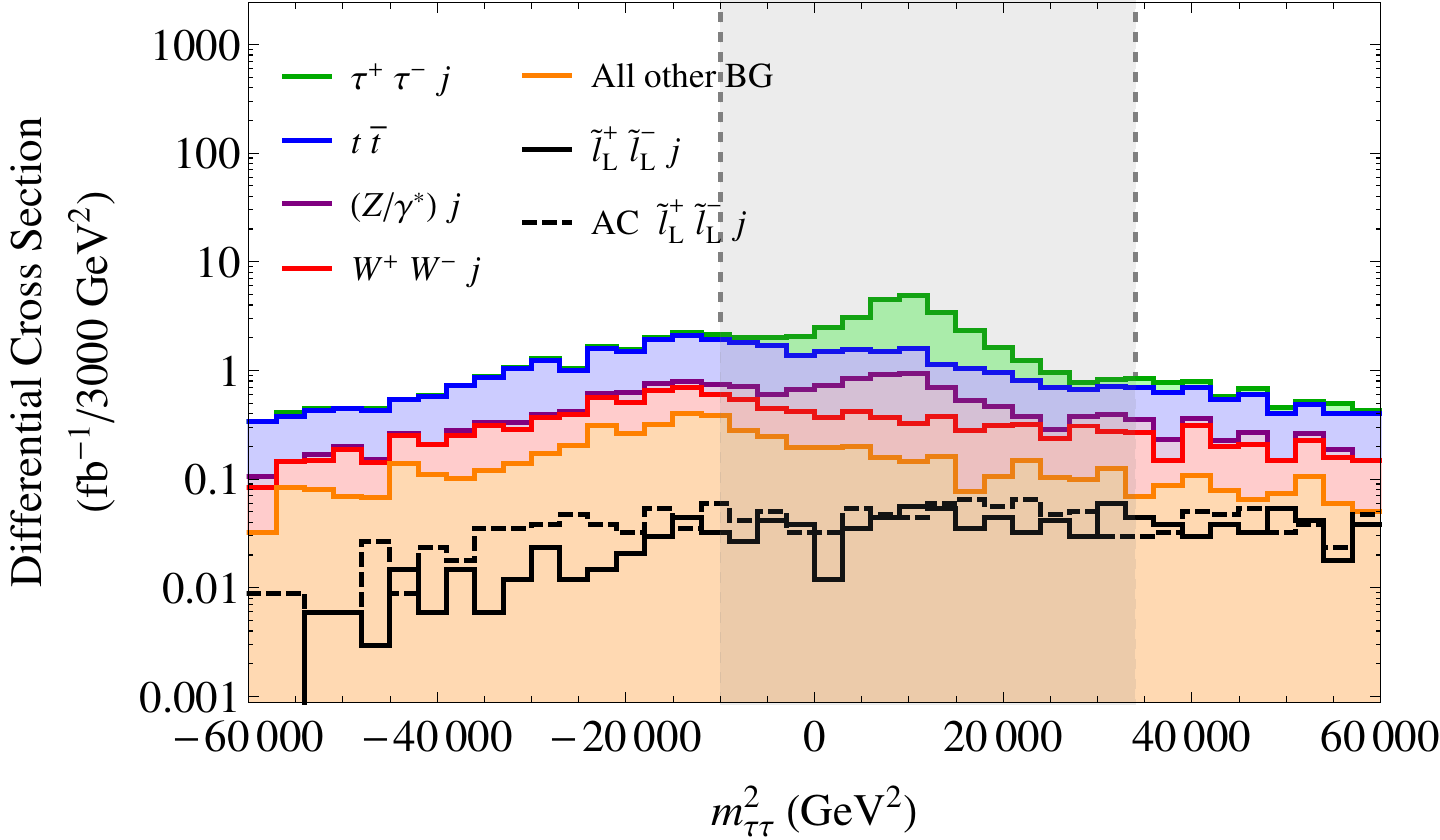}
  \caption{Reconstructed $m_{\tau \tau}^2$ distributions after cuts \ref{step:no b}-\ref{step:2 lep} for backgrounds (histograms stacked) and two different signal scenarios: degenerate left-handed selectrons and smuons with mass $m_{\tilde{l}_L} = 104$ GeV each decaying to a lepton plus either a 96 GeV neutralino LSP (solid black) or an auto-concealed modulino KK-tower (dashed black).  The gray area indicates the region removed in step \ref{step:mtautau} of the analysis.  This cut significantly reduces the $\tau \tau j$ and $Zj$ backgrounds.  }
  \label{fig:mtautau}
\end{figure}

\item \label{step:mtautau} Veto events with a reconstructed $m_{\tau \tau}^2$ between $-10,000~\text{GeV}^2 < m_{\tau \tau}^2 < 34,000~\text{GeV}^2$.  This is very effective at reducing the $\tau\tau j$ background and also helps pare down $Zj$.  The $m_{\tau \tau}^2$ distributions of backgrounds and representative signal models are shown in Fig. \ref{fig:mtautau}.

\begin{figure}[t]
  \centering
  \includegraphics[width=.7\columnwidth]{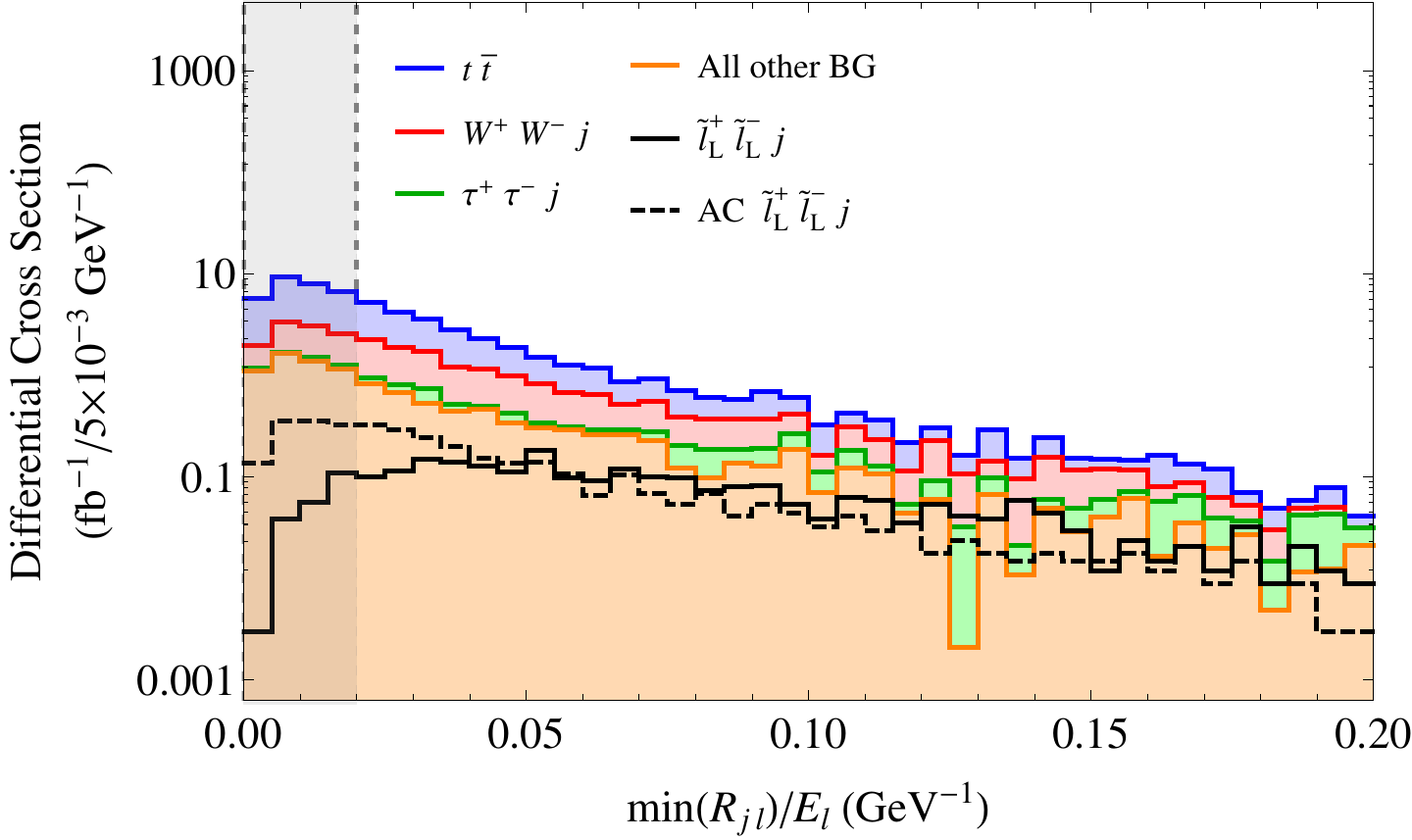}
  \caption{$\min(R_{jl})/E_{l}$ distributions after cuts \ref{step:no b}-\ref{step:mtautau} for backgrounds (histograms stacked) and two different signal scenarios: degenerate left-handed selectrons and smuons with mass $m_{\tilde{l}_L} = 104$ GeV each decaying to a lepton plus either a 96 GeV neutralino LSP (solid black) or an auto-concealed modulino KK-tower (dashed black).  The gray area indicates the region removed in step \ref{step:Rjl} of the analysis. }
  \label{fig:Rjl}
\end{figure}

\item \label{step:Rjl} Require $\min(R_{jl})/E_l > 0.02$ where $\min(R_{jl})$ gives the minimum distance $R$ between the hard jet and the leptons and $E_l$ is the energy of the lepton closest to the jet.  Cutting small $R_{jl}$ helps to isolate tops since their $b$-jets are closer to their leptons than the decay products from signal sleptons, which recoil against the hard ISR jet.  This is more effective than an invariant mass cut on $\min(m_{jl})$ since leptons from compressed spectra decays are softer than leptons from top/$W$ decays and this partially compensates for the closer distance between the jet and lepton.  We take advantage of this fact by dividing $R_{jl}$ by the lepton energy, further differentiating tops from signal while also cutting into the $WWj$ background. The $\min(R_{jl})/E_l$ distributions of backgrounds and representative signal models are shown in Fig. \ref{fig:Rjl}.

\begin{figure}[t]
  \centering
  \includegraphics[width=.7\columnwidth]{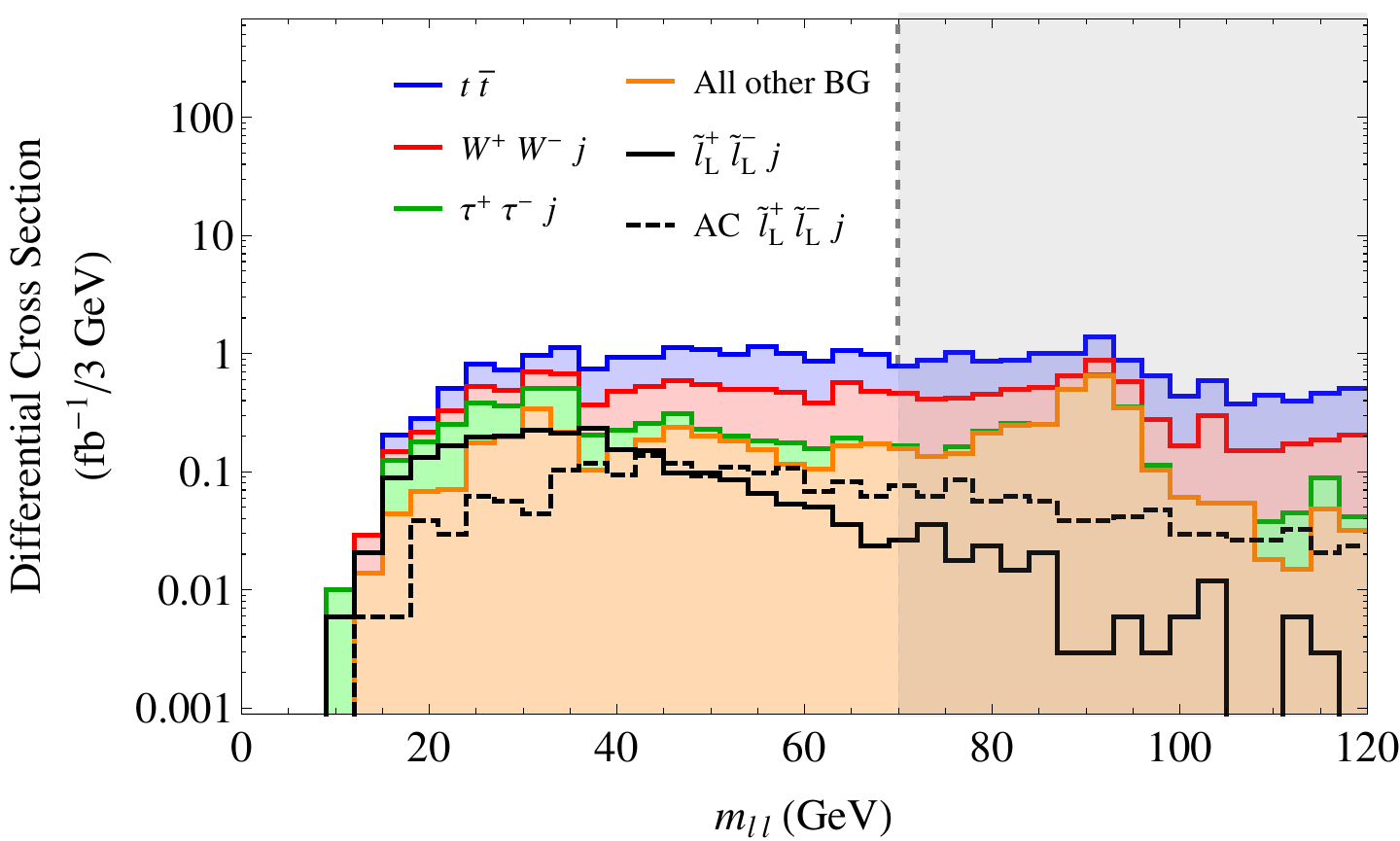}
  \caption{$m_{ll}$ distributions after cuts \ref{step:no b}-\ref{step:Rjl} for backgrounds (histograms stacked) and two different signal scenarios: degenerate left-handed selectrons and smuons with mass $m_{\tilde{l}_L} = 104$ GeV each decaying to a lepton plus either a 96 GeV neutralino LSP (solid black) or an auto-concealed modulino KK-tower (dashed black).  The gray area indicates the region removed in step \ref{step:mll} of the analysis.}
  \label{fig:mll}
\end{figure}

\item \label{step:mll} Require the two leptons' invariant mass $m_{ll}<70$ GeV.  This cuts away hard leptons from top/$W$ decays.  While cutting even lower on $m_{ll}$ can slightly extend the mass reach of the search, this comes at the expense of weakening the reach for larger $\Delta m$, since sleptons further separated from the LSP will have harder leptons. The $m_{ll}$ distributions of backgrounds and representative signal models are shown in Fig. \ref{fig:mll}.

\begin{figure}[t]
  \centering
  \includegraphics[width=.7\columnwidth]{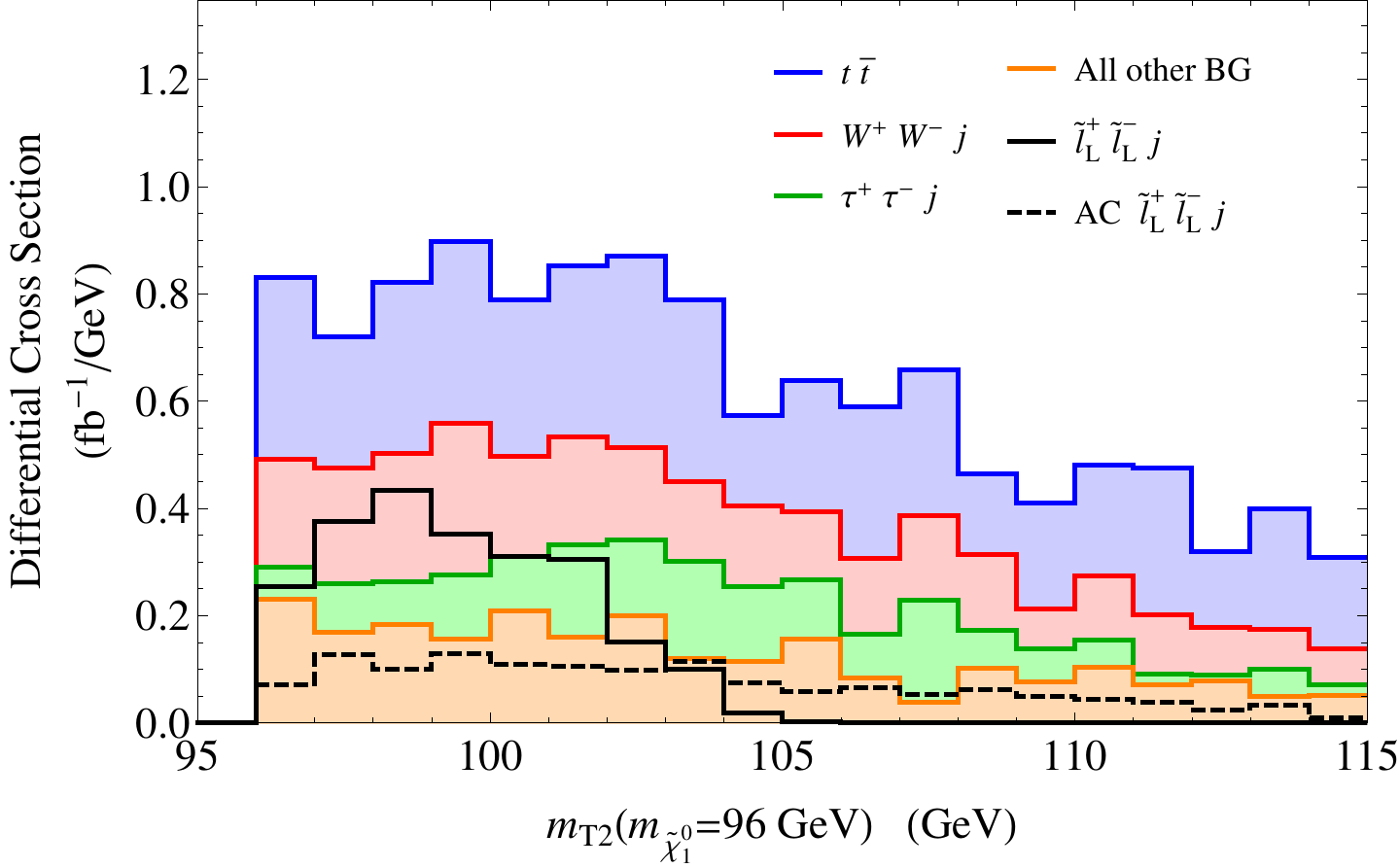}
  \caption{$m_{\T 2}(m_{\tilde{\chi}_1^0})$ distribution after cuts \ref{step:no b}-\ref{step:mll} for backgrounds  (histograms stacked) and signal using $m_{\tilde{\chi}_1^0}=96$ GeV for both.  Note that the vertical axis is not log-scaled.  Two different signal scenarios are shown: degenerate left-handed selectrons and smuons with mass $m_{\tilde{l}_L} = 104$ GeV each decaying to a lepton plus either a 96 GeV neutralino LSP (solid black) or an auto-concealed modulino KK-tower (dashed black).  When scanning the $(m_{\tilde{l}},m_{\tilde{\chi}_1^0})$ plane the expected signal and backgrounds at $(m_{\tilde{l}},m_{\tilde{\chi}_1^0})=(104,96)$ GeV  are evaluated in step \ref{step:mT2} using $m_{\T 2}(96~\text{GeV})<104~\text{GeV}$.  For the auto-concealed model, where multiple states play the role of LSP, we use three bins to evaluate the signal-- $m_{\T 2}(80~\text{GeV}),~m_{\T 2}(88~\text{GeV}),~m_{\T 2}(96~\text{GeV}) < 104~\text{GeV}$ --and choose the one with the largest signal significance.}
    \label{fig:mT2}
\end{figure}

\item \label{step:mT2} The final signal regions are defined using the stransverse mass as a function of the trial LSP mass $m_{\T 2}(\tilde{\chi}_1^0)$ as discussed in Sec.~\ref{sec: variables}, requiring $m_{\T 2}(\tilde{\chi}_1^0)$ is less than the trail $m_{\tilde l}$.  In this way, we scan the $m_{\tilde l}$ -- $m_{\tilde{\chi}_1^0}$ plane.  To evaluate the auto-concealed model, where multiple states play the role of LSP, we use three bins to evaluate the signal-- $m_{\T 2}(m_{\tilde{l}}-24~\text{GeV}),~m_{\T 2}(m_{\tilde{l}}-16~\text{GeV}),~m_{\T 2}(m_{\tilde{l}}-8~\text{GeV}) < m_{\tilde{l}}$ --and choose the one with the largest signal significance.  The $m_{\T 2}$ distribution for $m_{\tilde{\chi}_1^0} = 96$ GeV is shown in Fig. \ref{fig:mT2}.

\end{enumerate}

\begin{table}[h] \centering
\begin{tabular}{l|c|c|c|c|c|c|c|c||c|}
               & $\tau \tau$  & $t \bar t$   & $WWj$        & $ZZj$        & $WZj$        & $tW$         & $tq$         & $Zj$         & $l_L^+l_L^-j$ \\ \hline
MG cuts    & 3769013    & 5077488    & 74580      & 3478       & 21820      & 158000     & 2497581    & 4955268    & 10590    \\ \hline
1) $b$-veto         & 3167916    & 902842     & 61618      & 2883       & 18187      & 38262      & 795944     & 4240902    & 9302     \\ \hline
2) hard jet     & 417284     & 56546      & 19345      & 866        & 4548       & 8296       & 87754      & 1342713    & 1845     \\ \hline
3) $E_\T^{\rm{miss}}$  & 130135     & 25823      & 9855       & 478        & 1723       & 2784       & 46904      & 46654      & 1477     \\ \hline
4) SFOS $ll$      & 1751       & 4423       & 1726       & 186        & 289        & 553        & 169        & 570        & 327      \\ \hline
5) $m_{\tau \tau}^2$& 230      & 3511       & 1438       & 144        & 229        & 443        & 141        & 209        & 269      \\ \hline
6) $R_{jl_1}/E_l$ & 196        & 1613       & 905        & 75         & 141        & 171        & 137        & 115        & 248      \\ \hline
7) $m_{ll}$       & 174        & 713        & 397        & 8          & 25         & 85         & 113        & 60         & 229      \\ \hline \hline
8) $m_{\T2}$        & 94         & 256        & 164        & 3          & 8          & 32         & 67         & 30         & 227      \\ \hline
\end{tabular}
\caption{Background and signal counts for $pp$ collisions with $\sqrt{s}=14$ TeV and an integrated luminosity of $100~\rm{fb}^{-1}$.  This table shows the number of events which pass the Madgraph (MG) generator level cuts described in Sec. \ref{sec:bg} and analysis cuts \ref{step:no b}-\ref{step:mT2}.  The $m_{\T2}$ cut in this instance requires that $m_{\T2}(96~\rm{GeV})<104$ GeV. This corresponds to the signal model  $(m_{\tilde{l}},m_{\tilde{\chi}_1^0})=(104,96)~\text{GeV}$.   The far right column shows the cutflow for the matching model with degenerate left-handed selectrons and smuons with mass $m_{\tilde{l}_L} = 104$ GeV decaying to a SFOS lepton pair plus two 96 GeV neutralinos.}
\label{tab:cutflow}
\end{table}

Cutflows for the background and representative signal samples are show in Table \ref{tab:cutflow}.  Unfortunately, even after cuts \ref{step:no b}-\ref{step:mT2} a significant number of background events from $t \bar t$, $WWj$, and $\tau\tau j$ remain, as can be seen in Fig.~\ref{fig:mT2}. Since we will evaluate the signal significance as $s =S/\sigma_B$ where $S$ is the expected number of signal events and $\sigma_B$ is the total background uncertainty, even a moderate $\sigma_B$ $\left(\mathcal{O}(20 \%)\right)$ will overwhelm the signal.  If we use
\begin{equation}
\label{eq: sigma_B uncanceled}
\sigma_B=\sqrt{B+\sigma_{B\,\rm{sys}}^2}
\end{equation}
where $\sigma_{B\,\rm{sys}}$ is the systematic uncertainty of the background, and $B$ is the expected number of background events,\footnote{Here and in what follows we will use the Gaussian limit to approximate Poisson statistical errors as $\sigma_{B\,\rm{stat}}=\sqrt{B}$.} then it is clearly necessary to have good control over background systematics to obtain a detectable signal significance.  We attack this problem in the next section.

\subsection{Controlling background systematics: SF-DF}
\label{sec:SF-DF}

To reduce background systematics we estimate the size of the expected background using a different flavor (DF) di-lepton control region containing $e + \mu$ events which pass the equivalent of cuts \ref{step:no b}-\ref{step:mT2}.  Since sleptons are produced in same flavor (SF) opposite sign pairs only, while the dominant backgrounds produce different flavor lepton pairs as often as same flavor pairs, the DF control sample can be used to estimate--and hence subtract--the majority of the remaining backgrounds.

\begin{figure}[t]
  \centering
  \includegraphics[width=.7\columnwidth]{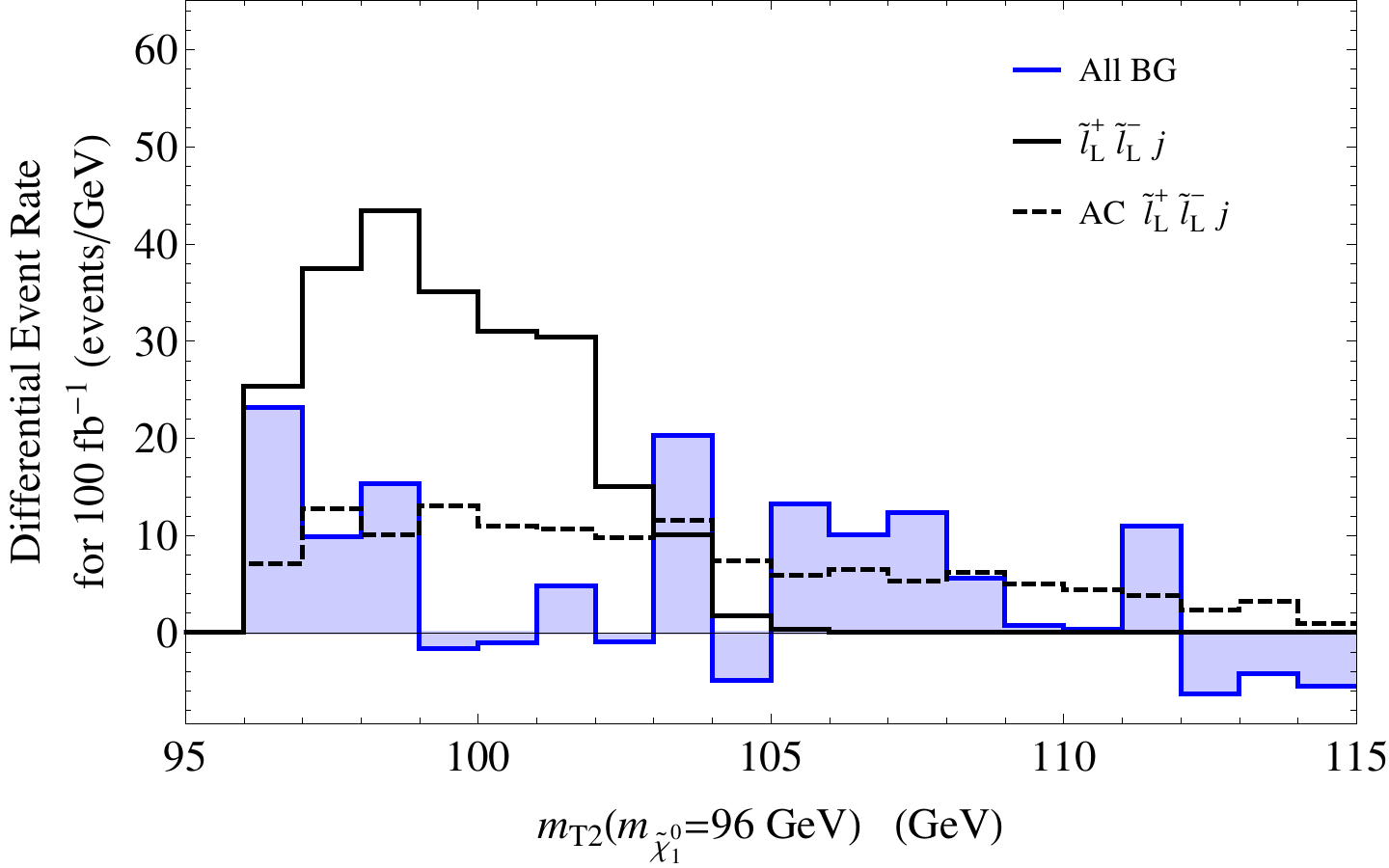}
  \caption{Different flavor subtracted $m_{\T 2}(p_\T^{l_1},p_\T^{l_2};m_{\tilde{\chi}_1^0})$ distribution after cuts \ref{step:no b}-\ref{step:mll} for backgrounds and signal  using $m_{\tilde{\chi}_1^0}=96$ GeV for both. Two different signal scenarios are shown: degenerate left-handed selectrons and smuons with mass $m_{\tilde{l}_L} = 104$ GeV each decaying to a lepton plus either a 96 GeV neutralino LSP (solid black) or an auto-concealed modulino KK-tower (dashed black).  Backgrounds are subtracted as described in Sec.~\ref{sec:SF-DF}.  The results are shown in this figure, not including the $m_{\T 2}$ cut described in step \ref{step:mT2}.}
    \label{fig:mT2BS}
\end{figure}

To do so we use a background subtraction scheme similar to that used in a CMS BSM search \cite{CMS:2014jfa}. Inside each signal region, we multiply the number of events in the DF control region $B_{\rm{DF}}$ by a normalization factor $n$ to account for the different efficiencies in detecting muons and electrons.  We then subtract the number of normalized control region events from the number of background signal region events $B_{\rm{SF}}$ to obtain the final background count: $B=B_{\rm{SF}}-n\,B_{\rm{DF}}$.  The residual systematic uncertainty in the corresponding number of background events should now come from uncertainty in $n$, which was determined to within $4\%$ in the CMS study.  We therefore take this as an estimate of our systematic uncertainty.  This comes at the cost of inflating the statistical uncertainty since the total background uncertainty is now

\begin{eqnarray}
\label{eq: sigma_B canceled}
\sigma_{B}  & = & 
\sqrt{ 
\left(\frac{\partial B}{\partial B_{\rm{SF}}}\right)^2 \sigma_{\rm{SF \, stat}}^2 + 
\left(\frac{\partial B}{\partial B_{\rm{DF}}}\right)^2 \sigma_{\rm{DF \, stat}}^2 +
\left(\frac{\partial B}{\partial n}\right)^2 \sigma_n^2
} \nonumber \\ 
	& = & \sqrt{B_{\rm{SF}} + n^2 B_{\rm{DF}} + B_{\rm{DF}}^2 \sigma_n^2}
\end{eqnarray}
where $\sigma_{\rm{SF\,stat}}$, and $\sigma_{\rm{DF\,stat}}$ are the statistical uncertainties in $B_{\rm{SF}}$ and $B_{\rm{DF}}$ respectively and $\sigma_n = 0.04 n$ is the total uncertainty in $n$ assumed in this study.   Despite the increased statistical error in this expression as compared to its non-background subtracted counterpart eq.~(\ref{eq: sigma_B uncanceled}), the lower systematic error more than compensates for this and reduces the total uncertainty.  For example, if we estimate $\sigma_{B\rm{\,sys}} = 0.2 B_{\rm{SF}}$ in (\ref{eq: sigma_B uncanceled}), then for the typical backgrounds present in our study the background subtracted expression for $\sigma_B$ in eq.~(\ref{eq: sigma_B canceled}) reduces the total error by more than $60\%$.    

To summarize, we define our signal significance as $s=S/\sigma_B$ with $B=B_{\rm{SF}}-n B_{\rm{DF}}$ and $\sigma_B$ defined in eq.~(\ref{eq: sigma_B canceled}).  An example of the signal as compared to the remaining background is shown in Fig.~\ref{fig:mT2BS}. 

\section{LHC14 reach with $100~\text{fb}^{-1}$}
\label{sec:reach}

\begin{figure}[t]
  \centering
  \includegraphics[width=.7\columnwidth]{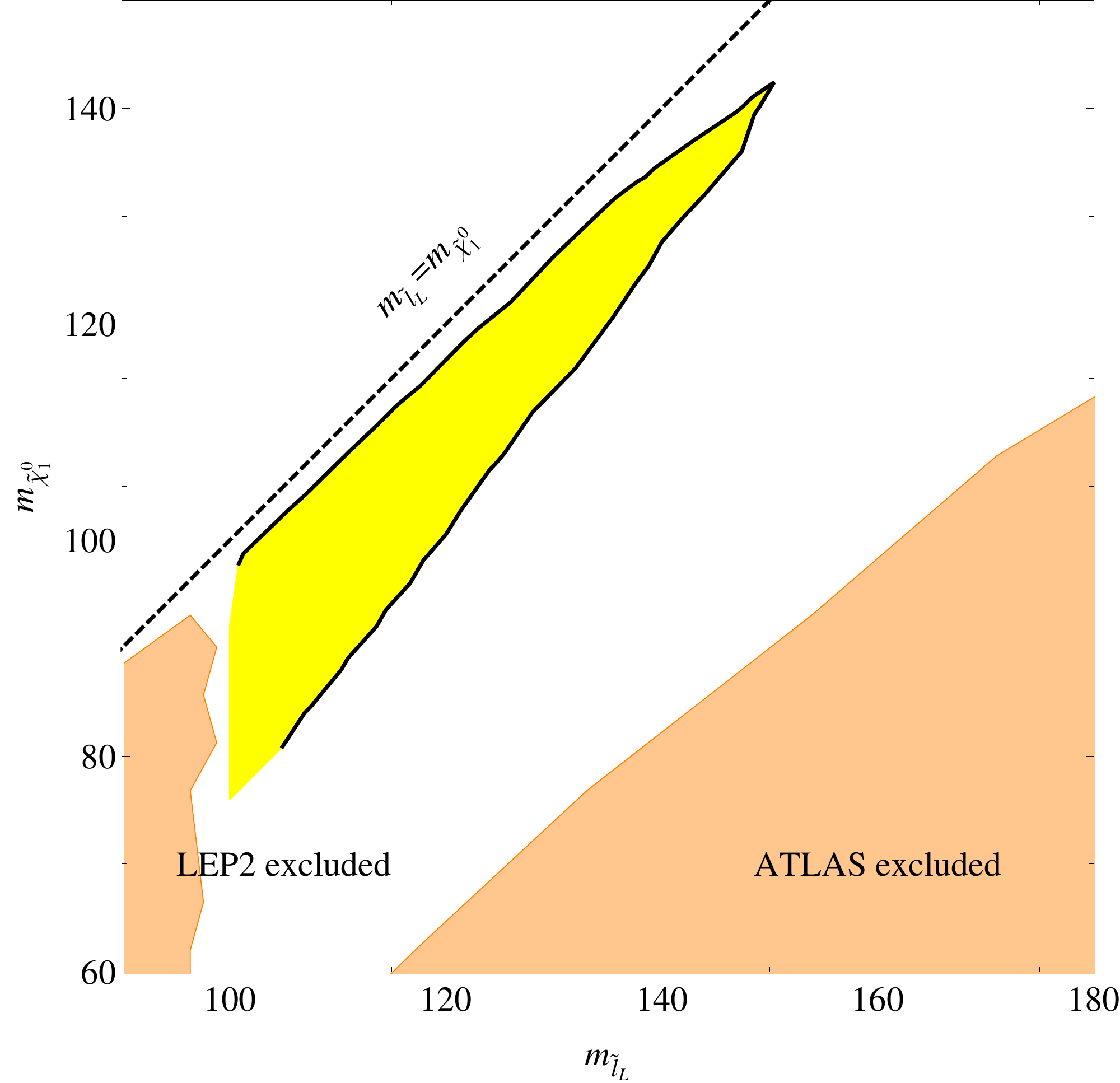}
  \caption{Potential $95\%$ CL exclusion reach at LHC14 with $100~\text{fb}^{-1}$ for degenerate left-handed smuons and selectrons.  Yellow area indicates the region excluded by this analysis, while orange areas indicate regions already excluded by LEP2 \cite{lepsusy} and ATLAS \cite{Aad:2014vma}.}
    \label{fig:LHreach}
\end{figure}

To evaluate the potential $95 \%$ confidence level (CL) exclusion reach of this analysis we demand a signal significance of 1.96.  As Fig.~\ref{fig:LHreach} shows, LHC14 with $100~\text{fb}^{-1}$ of data will be sensitive to sleptons with  $3~\text{GeV} < \Delta m < 24~\text{GeV} $ up to nearly $m_{\tilde l_L} \simeq 150$ GeV. As can be seen from the figure, this analysis will allow the LHC to explore beyond $m_{\tilde l_L} > 100$ GeV in the compressed region for the first time, though it will not completely close the gap between the compressed region and current ATLAS limits ($\Delta m \gtrsim 60$ GeV). On the other hand, the production cross sections for right-handed sleptons are such that this analysis will only reach $m_{\tilde l_R} = 100$ GeV with $100~\text{fb}^{-1}$, and therefore will not improve on LEP2 limits; at least until the LHC has more data.

The analysis is constrained to work in the region $\Delta m \lesssim 24$ GeV because it relies on $m_{\T 2}$ to concentrate the signal events into a narrow mass range.  The closer $m_{\tilde l}$ is to $m_{\tilde{\chi}_1^0}$ the more concentrated the signal events become and the better chance they have of beating the backgrounds.  However this only works down to about $\Delta m \simeq 3$ GeV because below this fewer leptons have enough energy to be tagged by the detector--even considering the fact that their parent sleptons are boosted from jet recoil.

\begin{figure}[t]
 \begin{minipage}{0.48\linewidth}
  \centering
  \includegraphics[width=.9\columnwidth]{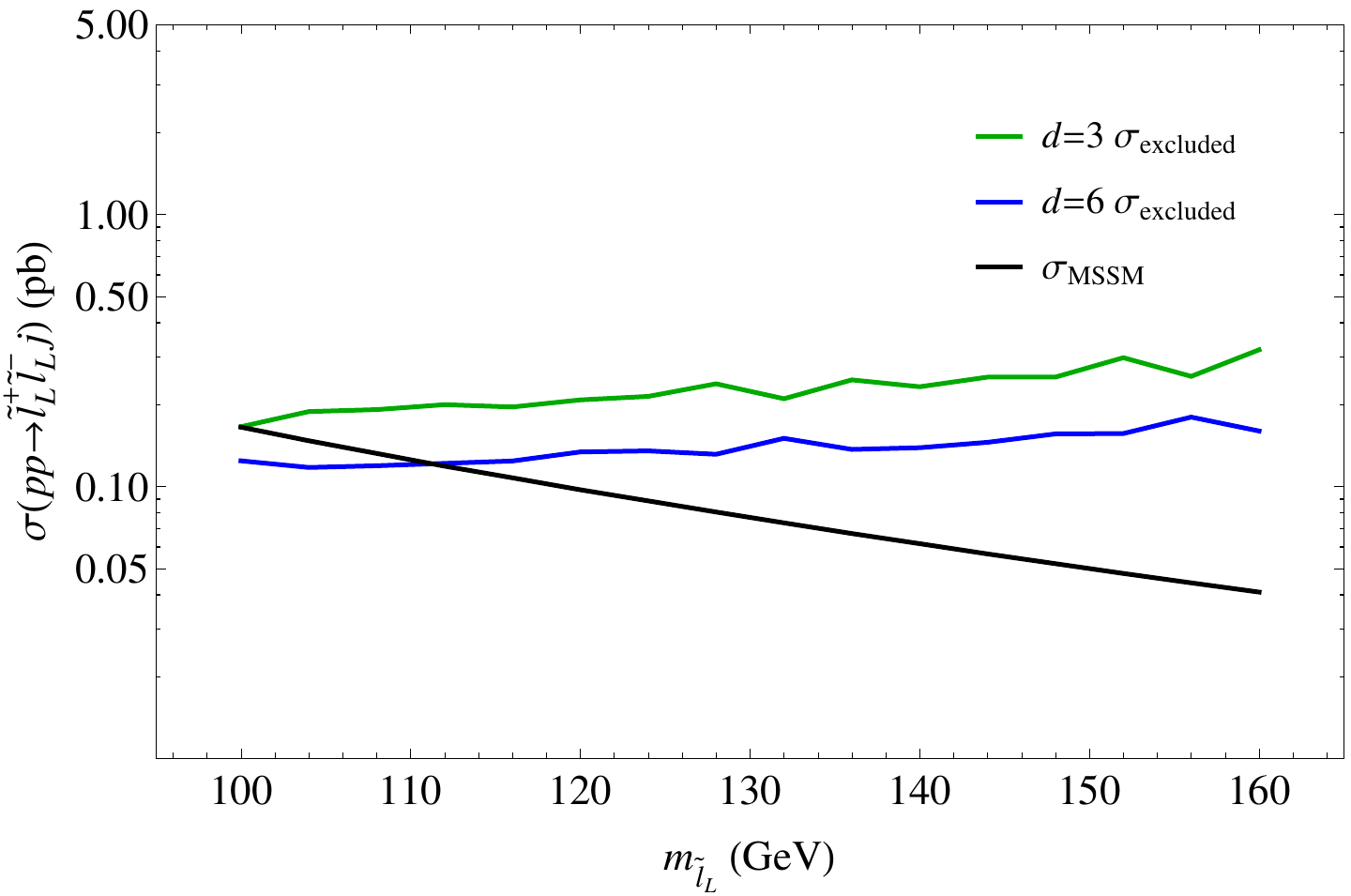}
   \end{minipage}
  \hspace{0.4 cm}
  \begin{minipage}{0.48\linewidth}
 \includegraphics[width=.9\columnwidth]{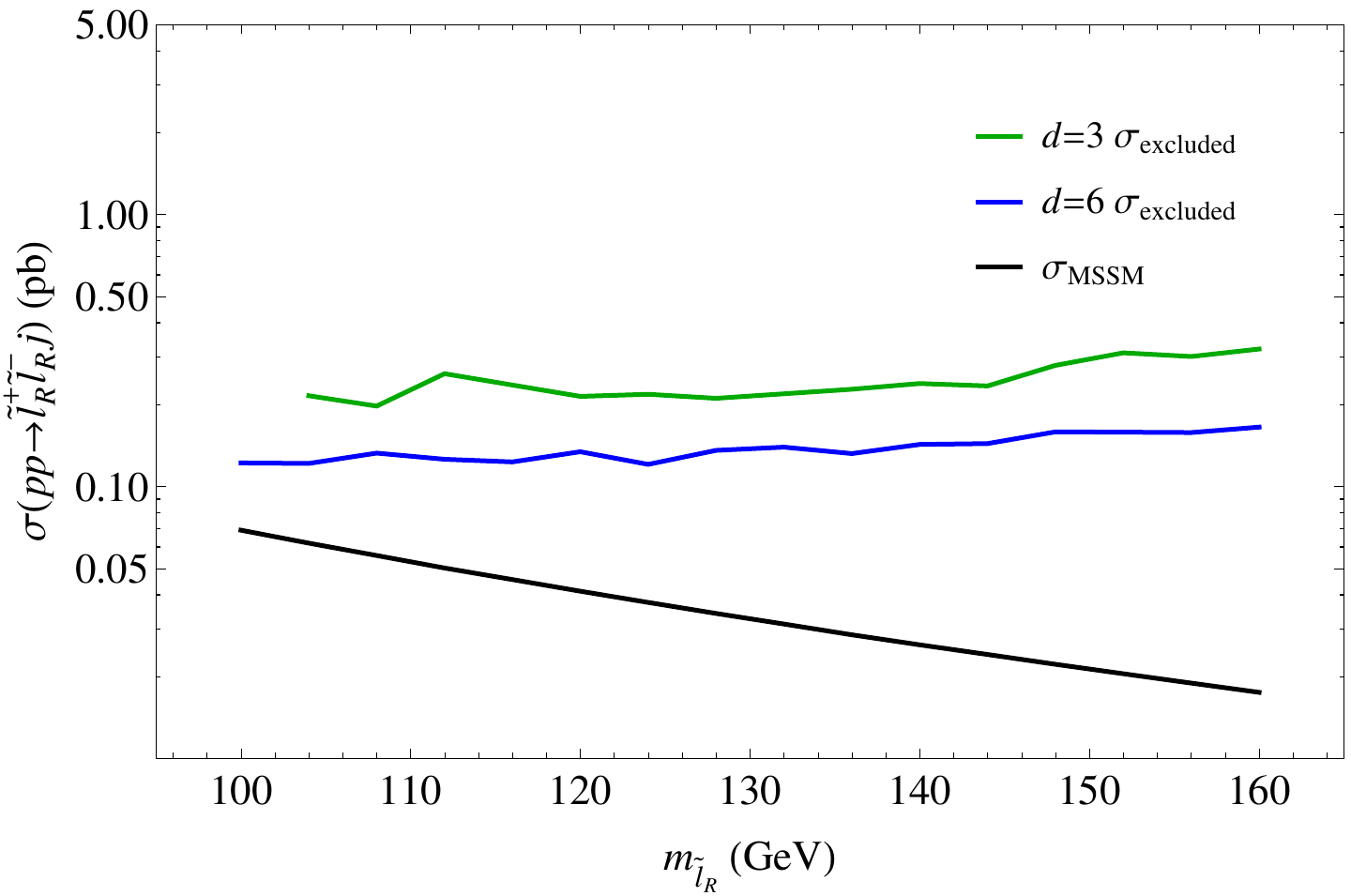}
 \end{minipage}
  \caption{Potential $95\%$ CL exclusion reach at LHC14 with $100~\text{fb}^{-1}$ for the production cross section of auto-concealed sleptons decaying to a modulino KK-tower in $d=3$ (green) or $d=6$ (blue) extra dimensions for left-handed (left plot) or right-handed (right plot) sleptons.  The black line shows the tree level production cross section for $pp\rightarrow \tilde{l}^+ \tilde{l}^- j$ indicating that, with this amount of data, the LHC should be sensitive to left-handed sleptons for $d=6~(3)$ up to 110 (100) GeV.  The production cross section of right-handed sleptons is smaller, requiring more data for sensitivity. }
    \label{fig:AClimits}
\end{figure}

In the case of auto-concealed SUSY detection becomes more challenging.  Figure \ref{fig:AClimits} shows the expected LHC14 $100~\text{fb}^{-1}$ reach for excluding the production cross section of AC sleptons decaying to the KK-tower of a modulino LSP which lives in either $d=6$ or 3 extra dimensions.  It will be able to exclude left-handed AC sleptons in $d=6~(3)$ up to 110 (100) GeV.  Again it will not be sensitive to right-handed AC sleptons except where LEP2 should have already been able to discover them.  

Clearly, AC sleptons present a bigger detection challenge than a normal compressed spectrum.  The reason for this is twofold.  First, the largest branching ratios are to particles with masses $\sim 80\%$ and $\sim 70\%$ of the slepton mass for $d=6$ and 3 respectively (see Fig. \ref{fig:auto-concealed BR}).  For 100 GeV left-handed sleptons this puts $\Delta m$ just inside the reach of the analysis in the first case, and just outside in the second.  As $m_{\tilde{l}}$ increases the effective $\Delta m$ only gets larger taking the model outside of the detection range.  The second challenge is that the differential branching ratio distributions are quite broad.  Therefore $m_{\T2}$ is not able concentrate signal events into a well defined window starting at the LSP mass.

Despite these challenges, it is encouraging to see that for the auto-concealed scenarios the sensitivity of the analysis improves as the number $d$ of extra dimensions increases, in contrast to what was observed in ref.~\cite{Dimopoulos:2014psa}, where higher $d$ cases were more difficult. Higher integrated luminosities and/or reduced systematic uncertainties would further increase sensitivity.

\section{Conclusions}
\label{sec:conclusions}

We have presented an analysis built to detect sleptons with small $m_{\tilde{l}}-m_{\tilde{\chi}_1^0}$. We use the $p p \rightarrow \tilde{l}^+ \tilde{l}^- j \rightarrow l^+ l^- \tilde{\chi}_1^0 \tilde{\chi}_1^0 j$ channel, where the neutral, collider stable $\tilde{\chi}_1^0$s are registered as $E_\T^{\rm{miss}}$ and the jet (which is required to have $p_\T >100$ GeV) comes from initial state radiation.  The jet requirement distinguishes this search from `classic' slepton analyses but is similar to monojet searches, and recent Higgsino search proposals \cite{Baer:2014kya,Han:2014kaa}. The jet requirement selects events with initial state radiation, and thus boosted sleptons. The resulting events tend to have large $E_\T^{\rm{miss}}$ and daughter leptons with $p_\T$ above the detection threshold.

After a series of cuts to remove BG events a significant number of unwanted backgrounds remain, most notably from the leptonic decays of $t \bar{t}$ and $W^+W^-j$.  Two final steps help to distinguish the signal from these processes.  First, we take advantage of the `stransverse' mass $m_{\T 2}$ to concentrate signal events into a window between the trial $m_{\tilde{l}}$ and $m_{\tilde{\chi}_1^0}$.  Finally, we use the fact that signal leptons come in SFOS pairs while the major remaining backgrounds produce different flavor pairs as often as same flavor pairs.  We use the different flavor pairs to `subtract away' the majority of the remaining background.  Though this introduces additional statistical error, this data-driven technique significantly reduces the systematic uncertainty and makes the signal detectable. 

The analysis presented in this paper should allow LHC14 with $100~\text{fb}^{-1}$ to search for degenerate left-handed selectrons and smuons in the compressed region $3~\text{GeV} < m_{\tilde{l}_L} - m_{\tilde{\chi}_1^0} < 24~\text{GeV}$ for $m_{\tilde{l}_L} \lesssim 150$ GeV.  This area is beyond LEP2 limits and currently unexplored by the LHC.  In addition, it should be sensitive to the challenging case of auto-concealed left-handed sleptons decaying to the KK-tower of a modulino LSP which lives in $d=6$ extra dimensions up to $m_{\tilde{l}_L} \lesssim 110$ GeV.  In both the compressed spectrum and auto-concealed scenarios this analysis will need more data to improve on LEP2 limits for right-handed sleptons. 

\subsection*{Acknowledgements}
We thank John March-Russell for the many useful discussions about theoretical considerations.  We thank  Mireia Crispin-Ortuzar, William Kalderon, William Fawcett, Claire Gwenlan, Koichi Nagai, and Jonathan Burr for their very helpful comments on this paper.  We thank Juan Rojo, Ulrich Haisch, and Emanuele Re for useful discussions regarding simulations and backgrounds.  We thank Jonathan Patterson for helping us with the Oxford Theory Computing Cluster.  AJB gratefully acknowledges the support of UK Science and Technology Facilities Council, the IPPP (Durham), and Merton College, Oxford.  JS gratefully acknowledges support from the United States Air Force Institute of Technology. The views expressed in this letter are those of the authors and do not reflect the official policy or position of the United States Air Force, Department of Defense, or the US Government.

\footnotesize
\bibliography{find_sleptons}
\end{document}